%% file: main.tex
\documentclass[sigconf,9pt]{cidr-2024}

\usepackage{graphicx}
\usepackage{balance}  
\usepackage{epsfig}
\usepackage{graphicx}
\usepackage{epstopdf}
\usepackage{latexsym}
\PassOptionsToPackage{noend}{algpseudocode}
\usepackage{algpseudocode}
\usepackage{pifont}
\usepackage{epsfig}
\usepackage{amssymb}
\usepackage{amsmath}
\usepackage{amsfonts}
\usepackage{subfigure}
\usepackage{stmaryrd}
\usepackage{url}
\usepackage{multirow}
\usepackage{array}
\usepackage[normalem]{ulem}
\usepackage{color}
\usepackage{cleveref}
\usepackage{xspace}
\usepackage{mathtools}
\usepackage{soul}
\usepackage{listings}
\usepackage{enumitem}
\usepackage{xcolor}
\usepackage{tikz}
\newsavebox{\blackball}
\newsavebox{\greenball}

\usepackage[utf8]{inputenc}
\usepackage{enumitem}
\usepackage{amsmath}

\usepackage{bbding}
\usepackage{amssymb}
\usepackage{epstopdf}
\usepackage{epsfig,endnotes}
\usepackage{url}
\usepackage{array}
\usepackage{booktabs}
\usepackage{threeparttable}
\usepackage{tabulary}

\usepackage{CJKutf8}
\usepackage[most,breakable]{tcolorbox}
\usepackage{etoolbox}
\usepackage{fancyhdr}
\usepackage{lipsum}
\usepackage[fencedCode]{markdown}

\usepackage[lined,boxed,vlined,ruled,linesnumbered]{algorithm2e}

\usepackage{bm}

\theoremstyle{definition}

\newcommand{\revise}[1]{\textcolor{blue}{#1}}

\newcolumntype{M}[1]{>{\centering\arraybackslash}m{#1}}
\errorcontextlines\maxdimen

\newcommand{\oursys}{\textsc{D-Bot}\xspace}

\newcommand{\llm}{\textsc{LLM}\xspace}
\newcommand{\llms}{\textsc{LLMs}\xspace}

\newcommand{\languagemodels}{{large language models}\xspace}

\newcommand{\dm}{{DM}\xspace}

\newcommand{\hi}[1]{\vspace{.25em} \noindent {\bf #1}\xspace}

 \newcommand{\bfit}[1]{\textbf{\textit{#1}}}

\definecolor{codegreen}{rgb}{0,0.6,0}
\definecolor{codegray}{rgb}{0.5,0.5,0.5}
\definecolor{codepurple}{rgb}{0.58,0,0.82}
\definecolor{backcolour}{rgb}{0.95,0.95,0.92}

\lstdefinestyle{mystyle}{
	backgroundcolor=\color{backcolour},   
	commentstyle=\color{codegreen},
	keywordstyle=\color{magenta},
	numberstyle=\tiny\color{codegray},
	stringstyle=\color{codepurple},
	basicstyle=\ttfamily\footnotesize,
	breakatwhitespace=false,         
	breaklines=true,                 
	captionpos=b,                    
	keepspaces=true,                 
	numbers=left,                    
	numbersep=5pt,                  
	showspaces=false,                
	showstringspaces=false,
	showtabs=false,                  
	tabsize=2
}
\lstdefinestyle{jsonStyle}{
	basicstyle=\small\ttfamily,
	columns=fullflexible,
	showstringspaces=false,
	commentstyle=\color{codegreen}\upshape,
	stringstyle=\color{codegreen},
	morestring=[b]",
	moredelim=[s][\color{codepurple}]{\{}{\}},
	moredelim=[s][\color{codepurple}]{[}{]},
	moredelim=[l][\color{codepurple}]{:},
	moredelim=[l][\color{codepurple}]{,}
}

\begin{document}

\title{LLM As Database Administrator}
\title{LLM As DBA}

\author{Xuanhe Zhou}
\affiliation{%
	\institution{Tsinghua University}
	\city{Beijing}
	\country{China}
}
\email{zhouxuan19@mails.tsinghua.edu.cn}

\author{Guoliang Li}
\affiliation{%
	\institution{Tsinghua University}
	\city{Beijing}
	\country{China}
}
\email{liguoliang@tsinghua.edu.cn}

\author{Zhiyuan Liu}
\affiliation{%
  \institution{Tsinghua University}
  \city{Beijing}
  \country{China}
}
\email{liuzy@tsinghua.edu.cn}

\pagestyle{plain}
\pagenumbering{arabic}

\input{abstract}

\maketitle

\input{01-introduction}

\input{03-rule}

\input{02-overview}
\input{03-knowledge}

\input{04-prompt}
\input{04-tool}

\input{05-feedback}

\input{05-update}

\input{06-experiment}
\input{08-relatedwork}
\input{09-conclusion}


\bibliographystyle{ACM-Reference-Format}
\bibliography{tool}

\newpage
\onecolumn
\appendix
\input{appendix}

\end{document}

%% file: abstract.tex

\begin{abstract}
\begin{sloppypar}
Database administrators (DBAs) play a crucial role in managing, maintaining and optimizing a database system   to ensure data availability, performance, and reliability. However, it is hard and tedious for DBAs to manage a large number of database instances (e.g., millions of instances on the cloud databases). Recently \languagemodels (\llms) have shown great potential to understand valuable documents and accordingly generate reasonable answers. Thus, we propose \oursys, a \llm-based database administrator that can continuously acquire database maintenance experience from textual sources, and provide reasonable, well-founded, in-time diagnosis and optimization advice for target databases. This paper presents a revolutionary \llm-centric framework for database maintenance, including $(i)$ database maintenance knowledge detection from documents and tools, $(ii)$ tree of thought reasoning for root cause analysis, and $(iii)$ collaborative diagnosis among multiple \llms. Our preliminary experimental results that \oursys can efficiently and effectively diagnose the root causes and our code is available at \href{github.com/TsinghuaDatabaseGroup/DB-GPT}{github.com/TsinghuaDatabaseGroup/DB-GPT}.
\end{sloppypar}
\end{abstract}

%% file: 01-introduction.tex




\section{Introduction}
\label{sec:intro}

\noindent \bfit{Limitations of DBAs.} Currently, most companies still rely on  DBAs for database maintenance (\dm,  e.g., tuning, configuring, diagnosing, optimizing) to ensure high performance, availability and reliability of the databases.  However, there is {\it a significant gap between DBAs and \dm tasks}.  First, it takes a long time to train a DBA. There are numerous relevant documents (e.g., administrator guides), which can span over 10,000 pages for just one database product and consumes DBAs several years to partially grasp the skills by applying in real practice. Second, it is hard to obtain enough DBAs to manage a large number of database instances, e.g.  millions of instance on cloud databases. Third, a DBA may not provide in-time response in emergent cases (especially for correlated issues across multiple database modules) and cause great financial losses. 





\begin{sloppypar}
\noindent \bfit{Limitations of Database Tools.} Many database products are equipped with semi-automatic maintenance tools to relieve the pressure of human DBAs~\cite{DBLP:journals/pacmmod/HuangWZTL023,DBLP:conf/ccgrid/LuXL0NZYZSZP22,DBLP:conf/ipccc/LiuZSMYP20,DBLP:conf/sigmod/KalmeghBR19,DBLP:conf/icde/LiuYZGCGLWLTL22}. However, they have several limitations. First, they are built by empirical rules~\cite{DBLP:conf/sigmod/YoonNM16,DBLP:conf/cidr/DiasRSVW05} or small-scale ML models (e.g., classifiers~\cite{DBLP:journals/pvldb/MaYZWZJHLLQLCP20}), which have poor text processing capability and cannot utilize available documents to answer basic questions. Second, they cannot flexibly generalize to scenario changes. For empirical methods, it is tedious to manually update rules by newest versions of documents. And learned methods require costly model retraining and are not suitable for online maintenance. Third, they cannot reason the root cause of an anomaly like DBAs, such as looking up more system views based on the initial analysis results. This capability is vital to detect useful information in complex cases.
\end{sloppypar}

\begin{figure}[!t]
	\vspace{.5em}
	\centering
	\includegraphics[width=0.95\linewidth, trim={0 0 1cm 0},clip]{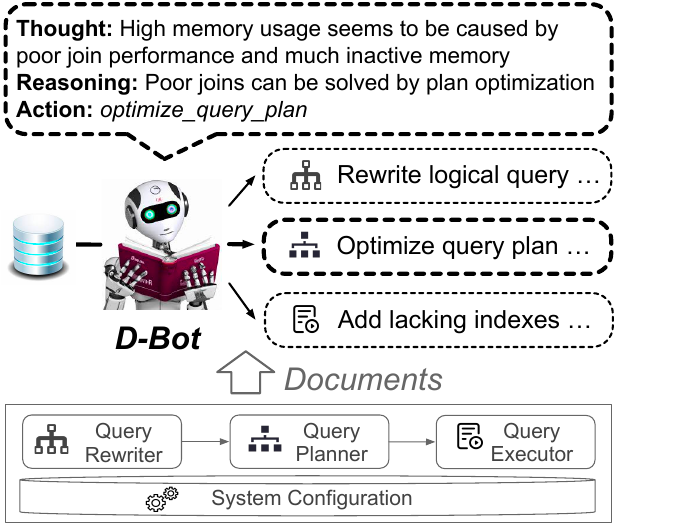}
	\vspace{.5em}
	\captionsetup{skip=0pt}
	\caption{\llm As DBA}
	\label{fig:example}
	\vspace{-1.5em}
\end{figure}

\noindent \bfit{Our Vision: A Human-Beyond Database Adminstrator.} To this end, we aim to build {\it a human-beyond ``DBA'' that can tirelessly learn from documents} (see Figure~\ref{fig:example}), which, given a set of  documents, automatically (1) learns experience from documents, (2) obtains status metrics by interacting with the database, (3) reasons about possible root causes with the abnormal metrics, and (4) accordingly gives optimization advice by calling proper tools.


\noindent \bfit{Challenges.} Recent advances in Large Language Models (\llms) have demonstrated superiority in understanding natural language, generating basic codes, and using external tools. However, leveraging \llm to design a ``human-beyond DBA'' is still challenging. 

{\it (1) Experience learning from documents.} Just like human learners taking notes in classes, although \llms have undergone training on vast corpus, important knowledge points (e.g., diagnosis experience) cannot be easily utilized without careful attention.  However, most texts are of long documents (with varying input lengths and section correlations) and different formats of the extracted experience can greatly affect the utilization capability of the \llm.


{\it (2)  Reasoning by interacting with database.} With the extracted experience, we need to inspire \llm to reason about the given anomalies. Different from basic prompt design in machine learning, database diagnosis is an interactive procedure with the database (e.g., looking up system views or metrics). However, \llm responses are often untrustworthy (``hallucination'' problem), and it is critical to design strategies that guide \llm to utilize proper interfaces of the database and derive reasonable analysis.




{\it (3) Mechanism for communication across multiple \llms.} Similar to human beings, one \llm alone may be stuck in sub-optimal solutions, and it is vital to derive a framework where multiple \llms collaborate to tackle complex database problems. By pooling their collective intelligence, these \llms can provide comprehensive and smart solutions that a single \llm or even skilled human DBA would struggle to think out.

\begin{figure*}[!t]
	\centering
	\includegraphics[width=\linewidth]{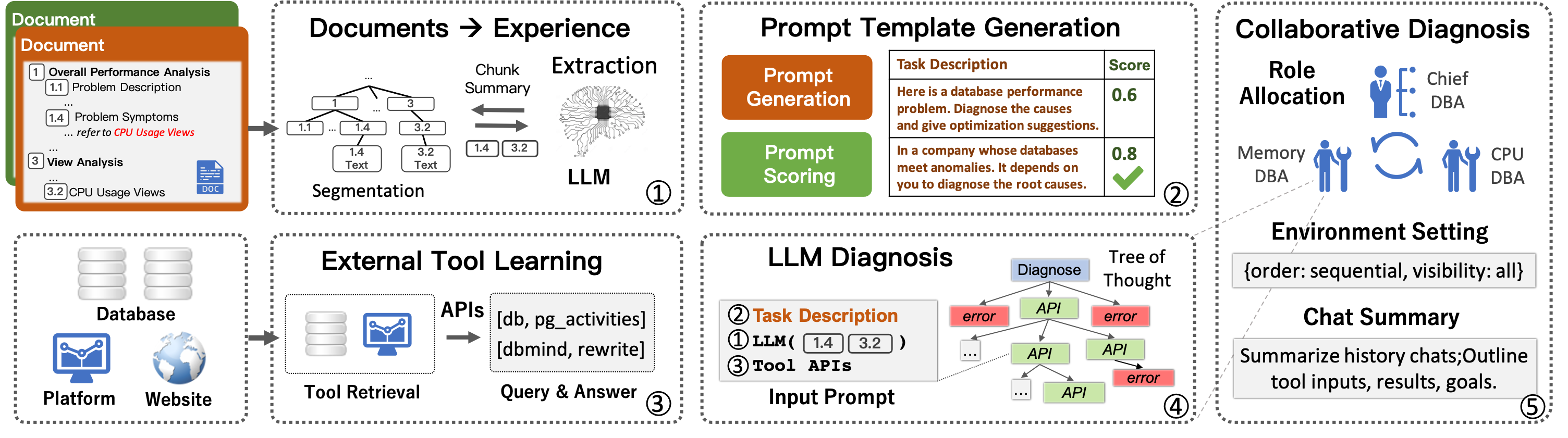}
	\vspace{-.75em}
	\captionsetup{skip=0pt}
	\caption{Overview of \oursys}
	\label{fig:overview}
	\vspace{-1em}
\end{figure*}

\noindent \bfit{Idea of \llm as DBA.} Based on above observations, we introduce \oursys, an LLM based database administrator. First, \oursys transforms documents into experiential knowledge by dividing them into manageable chunks and summarizing them for further extraction of maintenance insights with \llm. Second, it iteratively generates and assesses different formats of task descriptions to assist \llm in understanding the maintenance tasks better. Third, \oursys utilizes external tools by employing matching algorithms to select appropriate tools and providing \llm with instructions on how to use the APIs of selected tools. Once equipped with the experience, tools, and input prompt, \llm can detect anomalies, analyze root causes, and provide suggestions, following a {\it tree of thought} strategy to revert to previous steps if a failure occurs. {Moreover, \oursys promotes collaborative diagnosis by allowing multiple \llms to communicate based on predefined environmental settings, inspiring more robust solutions via debate-like communications. }

\hi{Contributions.} We make the following contributions.

\noindent (1) We design a \llm-centric database maintenance framework, and explore potential to overcome limitations of traditional strategies. 

\noindent (2) We propose an effective data collection mechanism by $(i)$ detecting experiential knowledge from documents and $(ii)$ leveraging external tools with matching algorithms. 

\noindent (3) We propose a root cause analysis method that utilizes \llm and tree search algorithm for accurate diagnosis. 

\noindent (4) We propose an innovative concept of collaborative diagnosis among \llms, thereby offering more comprehensive and robust solutions to complex database problems.

\noindent (5) Our preliminary experimental results that \oursys can efficiently and effectively diagnose the root causes.



%% file: 03-rule.tex

\section{Preliminaries}
\label{sec:preliminary}

\hi{Database Anomalies.} In databases, there are five common problems that can negatively affect the normal execution status. {\it (1) Running Slow.} The database exhibits longer response time than expectancy, leading to bad execution performance. {\it (2) Full Disk Capacity.} The database's disk space is exhausted, preventing it from storing new data. {\it (3) Execution Errors.} The database experiences errors, potentially due to improper error handling in the application (e.g., leaking sensitive data or system details) or issues within database (e.g., improper data types). {\it (4) Hanging.} The database becomes unresponsive, which is usually caused by long-running queries, deadlocks, or resource contention. {\it (5) Crashing.} The database unexpectedly shuts down, causing data inaccessible. {\it For a mature database product, each anomaly type is explained in the documentation and suitable to be learned by \llms.}

\hi{Observation Tools for Anomaly Detection.} ``Observability of the database'' is vital to detect above anomalies, including logs, metrics, and traces. \textit{(1) Logs} are records of database events. For example, PostgresSQL supports slow query logs (with error messages that can help debug and solve execution issues), but these logs may record a large scale of data and are generally not enabled in online stage. \textit{(2) Metrics} capture the aggregated database and system statistics. For example, views like pg\_stat\_statements record the templates and statistics of slow queries; tools like Prometheus~\cite{turnbull2018monitoring} provide numerous monitoring metrics, making it possible to capture the real time system status. \textit{(3) Traces}  provide visibility into how requests behave during executing in the database. Different from logs that help to identify the database problem, traces help to locate the specific abnormal workload or application. 


\hi{Optimization Tools for Anomaly Solving.} Users mainly concern how to restore to the normal status after an anomaly occurs. Here we showcase some optimization tools. (1) For slow queries, since most open-source databases are weak in logical transformation, there are external engines (e.g., Calcite with $\sim$120 query rewrite rules) and tuning guides (e.g., Oracle with over 34 transformation suggestions) that help to optimize slow queries. (2) For knob tuning, many failures (e.g., max\_connections in Postgres) or bad performance (e.g., memory management knobs) are correlated with database knobs (e.g., for a slow workload, incresae innodb\_buffer\_pool\_size in MySQL by 5\% if the memory usage is lower than 60\%). Similarly, there are index tuning rules that generate potentially useful indexes (e.g., taking columns within the same predicate as a composite index). Besides, we can utilize more advanced methods, such as selecting among heuristic methods~\cite{AutoAdmin,Drop,DB2Advis} and learned methods~\cite{zhou2020database,SWIRL,DBABandits,CIKM20,SmartIX,AutoIndex,BudgetMCTS} for problems like {\it index lacking}, which is not within the scope of this paper.

We aim to design \oursys, an LLM-based DBA, for automatically diagnosing the database anomalies and use LLM to directly  (or call appropriate tools to indirectly) provide the root causes.

%% file: 02-overview.tex
\vspace{-1em}

\section{The Vison of \oursys}
\label{sec:overview}


Existing \llms are criticized for problems like ``Brain in a Vat''~\cite{DBLP:journals/corr/abs-2307-03762}. Thus, it is essential to establish close connections between \llms and the target database, allowing us to guide \llms in effectively maintaining the database's health and functionality. Hence, we propose \oursys, which is composed of two stages. 


First, in preparation stage, \oursys generates experience (from documents) and prompt template (from diagnosis samples), which are vital to guide online maintenance. 


\begin{compactitem}
	\item
	\bfit{Documents $\rightarrow$ Experience.} Given a large volume of diverse, long, unstructured  database documents (e.g., database manual,  white paper, blogs), we first split each document into chunks that can be processed by the \llm. To aggregate correlated chunks together (e.g., chunk $v_i$ that explains the meaning of ``bloat-table'' and chunk $v_j$ that utilizes ``bloat-table'' in root cause analysis), we generate a summary for each chunk based on both its content and its subsections. Finally, we utilize \llm to extract maintenance experience from chunks with similar summaries (Section~\ref{sec:prepare}).
	
	\item
	\bfit{Prompt Template Generation.} To help \llm better understand the \dm tasks, we iteratively generate and score different formats of task descriptions using \dm samples (i.e., given the anomaly and solutions, ask \llm to describe the task), and adopt task description that both scores high performance and is sensible to human DBAs (in cases of learning bias) for \llm diagnosis (Section~\ref{sec:prompt}).
\end{compactitem}

Second, in maintenance stage, given an anomaly, \oursys iteratively reasons the possible root causes by taking advantages of external tools and multi-\llm communications.

\begin{compactitem}
	\item
	\bfit{External Tool Learning.} For a given anomaly, \oursys first matches relevant tools using algorithms like Dense Retrieval. Next, \oursys provides the tool APIs together with their descriptions to the \llm (e.g., function calls in GPT-4). After that, \llm can utilize these APIs to obtain metric values or optimization solutions. For example, in PostgresSQL, \llm can acquire the templates of slowest queries in the {\it pg\_activity} view. If these queries consume much CPU resource (e.g., over 80\%), they could be root causes and optimized with rewriting tool (Section~\ref{sec:tool}).
	
	
	\item
	\bfit{\llm Diagnosis.} Although \llm can understand the functions of tool APIs, it still may generate incorrect API requests, leading to diagnosis failures. To solve this problem, we employ the {\it tree of thought} strategy, where \llm can go back to previous steps if the current step fails. It significantly increases the likelihood of \llms arriving at reasonable diagnosis results (Section~\ref{sec:diagnose}).
	
	\item
	\bfit{Collaborative Diagnosis.} A single \llm may execute only the initial diagnosis steps and end up early,  leaving the problem inadequately resolved. To address this limitation, we propose the use of multiple \llms working collaboratively. Each \llm plays a specific role and communicates by the environment settings (e.g., priorities, speaking orders). In this way, we can enable \llms to engage in debates and inspire more robust solutions (Section~\ref{sec:collaborate}). 
	
\end{compactitem}

%% file: 03-knowledge.tex

\section{Experience Detection From Documents}
\label{sec:prepare}

Document learning aims to extract experience segments from textual sources, where the extracted segments are potentially useful in different \dm cases. For instance, when analyzing the root causes of performance degradation, \llm utilizes the {\it ``many\_dead\_tuples''} experience to decide whether dead tuples have negatively affected the efficiency of index lookup and scans.

\hi{Desired Experience Format.} To ensure \llm can efficiently utilize the experience, each experience fragment should include four fields. As shown in the following example, {\it ``name''} helps \llm to understand the overall function; {\it ``content''} explains how the root cause can affect the database performance (e.g.,  the performance hazards of many dead tuples); {\it ``metrics''} provide hints of matching with this experience segment, i.e., \llm will utilize this experience if the abnormal metrics exist in the ``metrics'' field; {\it ``steps''} provide the detailed procedure of checking whether the root cause exists by interacting with database (e.g., obtaining the ratio of dead tuples and live tuples from table statistics views).
\begin{lstlisting}
"name": "many_dead_tuples",
"content": "If the accessed table has too many dead tuples, it can cause bloat-table and degrade performance",
"metrics": ["live_tuples", "dead_tuples", "table_size", "dead_rate"],
"steps": "For each accessed table, if the total number of live tuples and dead tuples is within an acceptable limit (1000), and table size is not too big (50MB), it is not a root cause. Otherwise, if the dead rate also exceeds the threshold (0.02), it is considered a root cause. And we suggest to clean up dead tuples in time."
\end{lstlisting}


\hi{\llm for Experience Detection.} It aims to detect experience segments that follow above format. Since different paragraphs within a long document may be correlated with each other (e.g., the concept of ``bloat-table'' appearing in ``many\_dead\_tuples'' is introduced in another section), we explain how to extract experience segments without losing the technical details.

\textit{Step1: Segmentation.} Instead of partitioning documents into fixed-length segments, we divide them based on the structure of the section structures and their content. Initially, the document is divided into chunks using the section separators. If a chunk exceeds the maximum chunk size (e.g., 1k tokens), we further divide it recursively into smaller chunks.

\textit{Step2: Chunk Summary.} Next, for each chunk denoted as $x$, a summary $x.summary$ is created by feeding the content of $x$ into \llm with a summarization prompt $p_{summarize}$:

\begin{tcolorbox}[enhanced,sharp corners,
	width={8.5cm},
	colback=white,
	borderline={0.3mm}{0.3mm}{white},
	boxsep=-.5mm]
	\bfit{$p_{summarize}$} = {\it Summarize the provided chunk briefly $\cdots$ Your summary will serve as an index for others to find technical details related to database maintenance $\cdots$ Pay attention to examples even if the chunks covers other topics.}
\end{tcolorbox}

The generated $x.summary$ acts as a textual index of $x$, enabling the matching of chunks containing similar content.

\textit{Step3: Experience Extraction.} Once the summaries of the chunks are generated, \llm parses the content of each chunk and compares it with the summaries of other chunks having similar content, which is guided by the extraction prompt $p_{extract}$. This way, experience segments that correlate with the key points from the summaries are detected.

\begin{figure*}[!t]
	\centering
	\includegraphics[width=.98\linewidth]{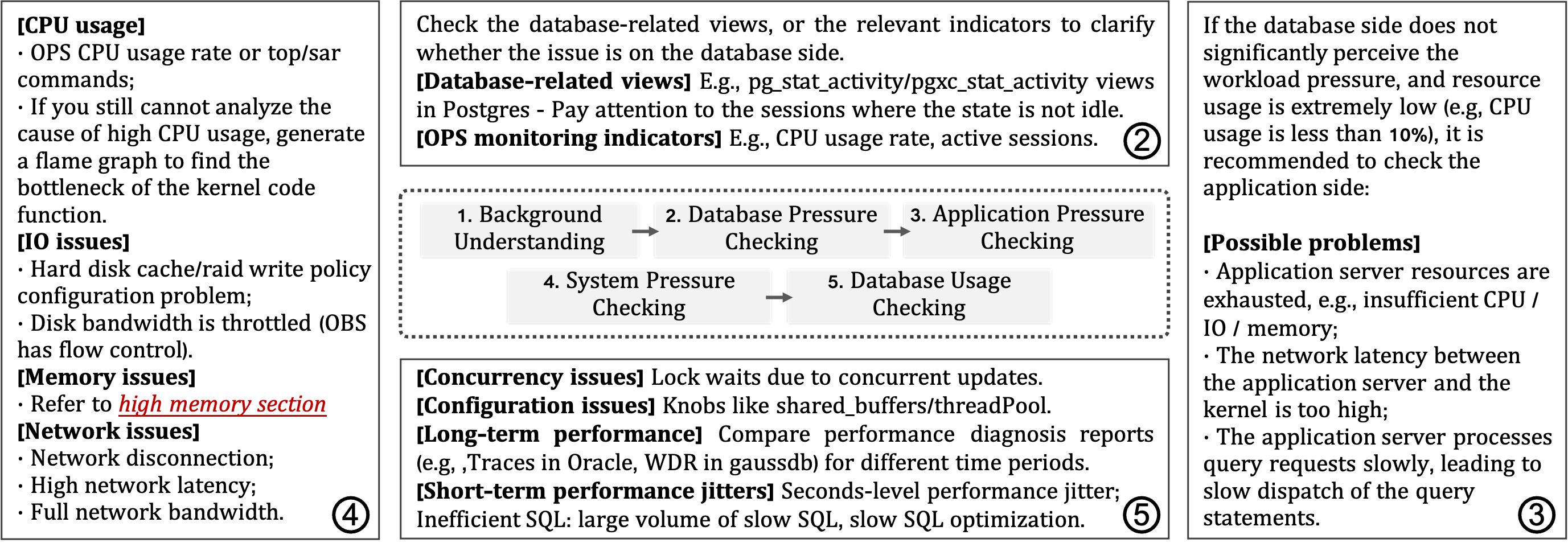}
	\vspace{.5em}
	\captionsetup{skip=0pt}
	\caption{The outline of diagnosis experience extracted from documents.}
	\label{fig:knowledge}
\end{figure*}

\begin{tcolorbox}[enhanced,sharp corners,
	width={8.5cm},
	colback=white,
	borderline={0.3mm}{0.3mm}{white},
	boxsep=-.5mm]
	\bfit{$p_{extract}$} = {\it Given a chunk summary, extract diagnosis experience from the chunk.} {\it If uncertain, explore diagnosis experience in chunks with similar summaries.}
\end{tcolorbox}

In our implementation, given a document, we use \llm to extract experience segments into the above 4-field format.




\hi{Detected Maintenance Experience.} In Figure~\ref{fig:knowledge}, we showcase the simplified diagnosis procedure together with some necessary details, coming from chunks originally in different sections of the given documents (e.g., the maintenance guide with over 100 pages).

{\tt 1. Background Understanding.} It's crucial to grasp the context of system performance, such as recent changes in customer expectation, workload type, or even system settings.

\begin{sloppypar}
{\tt 2. Database Pressure Checking.} This step identifies database bottlenecks, such as  tracking CPU usage and active sessions; and monitoring system views (e.g., \textit{pg\_stat\_activity} and \textit{pgxc\_stat\_activity}) to focus on non-idle sessions. 
\end{sloppypar}

{\tt 3. Application Pressure Checking.} If there is no apparent pressure on the database or the resource consumption is very low (e.g., CPU usage below 10\% and only a few active sessions), it is suggested to investigate the application side, such as exhausted application server resources, high network latency, or slow processing of queries by application servers.

{\tt 4. System Pressure Checking.} The focus shifts to examining the system resources where the database is located, including CPU usage, IO status, and memory consumption. 

{\tt 5. Database Usage Checking.} Lastly, we can investigate sub-optimal database usage behaviors, such as (1) addressing concurrency issues caused by locking waits, (2) examining database configurations, (3) identifying abnormal wait events (e.g., io\_event), (4) tackling long/short-term performance declines, and (5) optimizing poorly performing queries that may be causing bottlenecks.

%% file: 04-prompt.tex

\section{Diagnosis Prompt Generation}
\label{sec:prompt}

Instead of directly mapping extracted experience to new cases, next we explore how to teach \llms to (1) understand the database maintenance tasks and (2) reason over the root causes by itself.


\hi{Input Enrichment.} With a database anomaly $x$ as input, we can enrich $x$ with additional description information so called input prompt $x'$. On one hand, $x'$ helps \llm to better understand the task intent. On the other hand, since database diagnosis is generally a complex task that involves multiple steps, $x'$ preliminarily implies how to divide the complex task into sub-tasks in a proper order, such that further enhancing the reasoning of \llm.


From our observation, the quality of $x'$ can greatly impact the performance of \llm on maintenance tasks~\cite{Zhou2022a} (Figure~\ref{fig:overview}). Thus, we first utilize \llm to suggest candidate prompts based on a small set of input-output pairs (e.g., 5 pairs for a prompt). Second, we rank these generated prompts based on a customized scoring function (e.g., the ratio of detected root causes), and reserve the best prompts (e.g., top-10) as candidates. Finally, we select the best one to serve as the input prompt template for the incoming maintenance tasks.

%% file: 04-tool.tex

\section{External Tool Learning}
\label{sec:tool}


As we know, the efficient use of tools is a hallmark of human cognitive capabilities~\cite{qin2023tool,qin2023toolllm}. When human beings encounter a new tool, they start to understand the tool and explore how it works, i.e., taking it as something with particular functions and trying to understand what the functions are used for. Likewise, we aim to inspire similar ability within \llm.

\hi{Tool Retrieval.} We first retrieve the appropriate tools for the diagnosis task at hand, represented as $D_t$. There are several methods used, such as BM25, \llm Embeddings, and Dense Retrieval.  

{\it (1) BM25}, simply represented as $f(D_t, Q) = \text{BM25}$, is a common probabilistic retrieval method that ranks tool descriptions ($D$) based on their relevance to the given anomaly ($Q$)~\cite{robertson2009probabilistic}. 

{\it (2) \llm Embeddings}, denoted as $f(D_t, L) = {LLM}_E$, are a method that converts tool descriptions ($D_t$) into embeddings ($E_t$) using \llm, i.e., $E_t = L(D_t)$. These embeddings capture the semantic meanings in a multi-dimensional space, hence helping in finding related tools even in the absence of keyword overlap, $D_t = LLM_E(E_t)$. 

{\it (3) Dense Retrieval}, denoted as $f(Q, D_t, N) = D_R$, uses neural networks ($N$) to generate dense representations of both the anomaly ($Q$) and the tool descriptions ($D_t$), separately denoted as  $\text{Dense}_Q$ and $\text{Dense}_D$. To retrieve the relevant tools, we calculate the similarity between $\text{Dense}_Q$ and $\text{Dense}_D$, and rank them based on these similarity scores.

The proper method for tool retrieval depends on the specific scenarios. {\it BM25} is efficient for quick results with large volumes of API descriptions in the tools and clear anomaly characters. {\it \llm Embeddings} excel at capturing semantic and syntactic relationships, which is especially useful when relevance isn't obvious from keywords (e.g., different metrics with similar functions). {\it Dense Retrieval} is ideal for vague anomaly, which captures context and semantic meaning, but is more computational costly.

%% file: 05-feedback.tex
\begin{figure}[!t]
	\vspace{.5em}
	\centering
	\includegraphics[width=.98\linewidth, trim={0 0 0 0},clip]{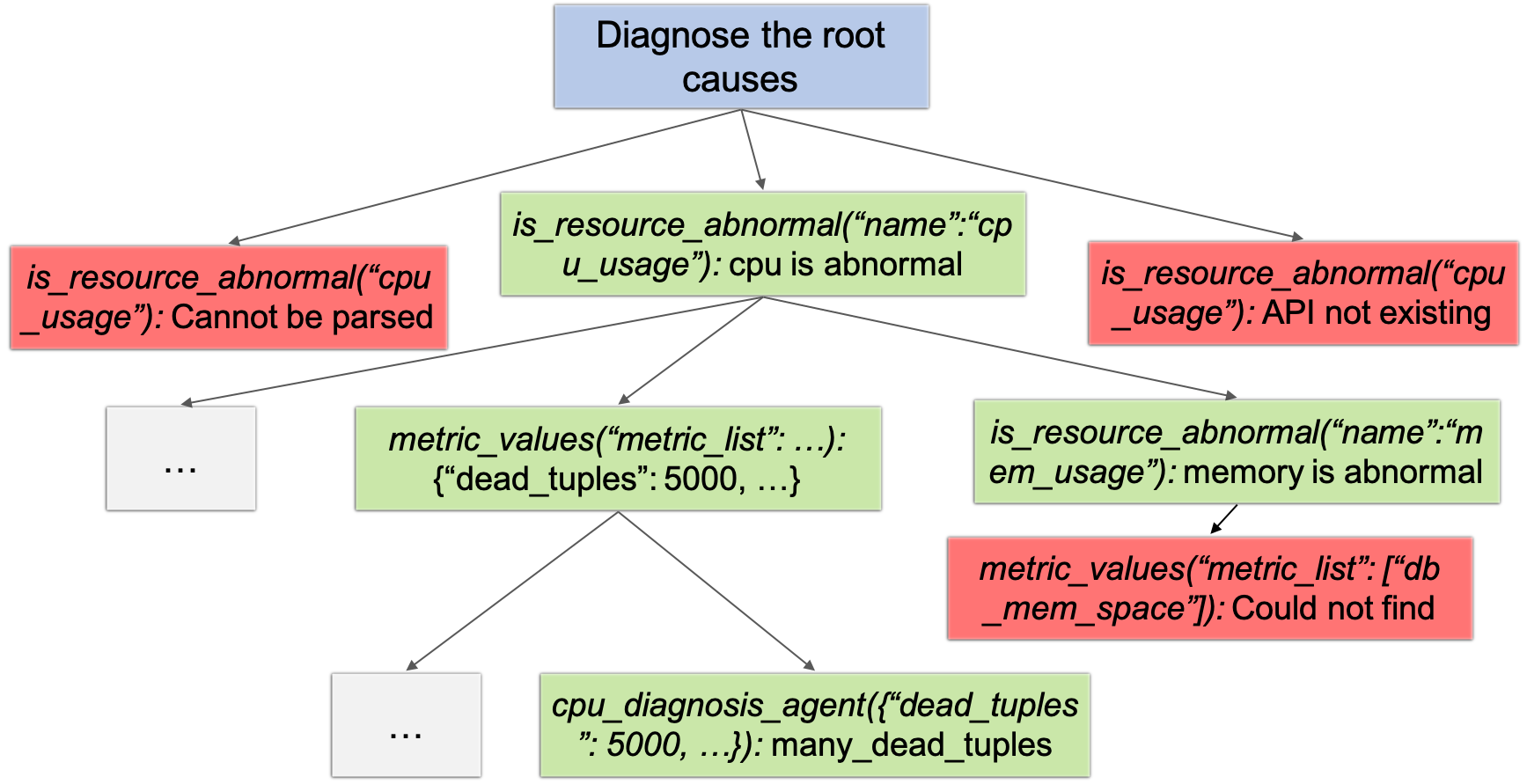}
	\vspace{-.5em}
	\caption{Example \llm diagnosis by tree of thought}
	\label{fig:tot}
	\vspace{-1.25em}
\end{figure}

\section{\llm Diagnosis}
\label{sec:diagnose}

\hi{Tree Search Algorithm using \llm.} To avoid diagnosis failures caused by the incorrect actions (e.g., non-existent API name) derived by \llm, we propose to utilize the {\it tree of thought} strategy that can guide \llm to go back to previous actions if the current action fails. 

\textit{Step1: Tree Structure Initialization.} We initialize a tree structure, where root node is the diagnosis request (Figure~\ref{fig:tot}). Utility methods are utilized to manipulate the tree structure, and UCT score for node $v$ are computed based on the modifications during planning, i.e., $\text{UCT}(v) = \frac{w(v)}{n(v)} + C \cdot \sqrt{\frac{\ln(N)}{n(v)}}$, where $\frac{\ln(N)}{n(v)}$ denotes the selection frequency and $w(v)$ denotes the success ratio of detecting root causes. Note, the action of $n(v$ fails to call tool API,  $w(v)$ equals -1.

\textit{Step2: Simulate Execution.} This step kickoffs the execution of simulations starting from the root node of the tree. It involves selecting nodes based on specific standard (e.g., detected abnormal metrics). If the criteria for selecting a new node is met, a new node is chosen; otherwise, the node with the highest UCT value is selected. 

\textit{Step3: Existing Node Reflection.} For each node in the path from the root node to the selected node, reflections are generated based on decisions made at previous nodes. For example, we count on \llm to rethink the benefits of analyzing non-resource relevant metrics. If \llm decides the action cannot find any useful information, the UCT value will be reduced and set to that of its parent node. In this way, we can enhance the diagnosis efficiency.

\textit{Step4: Terminal Condition.} If \llm cannot find any more root cause (corresponding to a leaf node) for a threshold time (e.g., five), the algorithm ends and \llm outputs the final analysis based on the detected root causes.

%% file: 05-update.tex
\section{Collaborative Diagnosis For Complex Cases}
\label{sec:collaborate}

 A single \llm may be limited in its ability to fully resolve a problem (e.g., stuck in initial steps). Collaborative diagnosis involves the utilization of multiple \llms to collectively address complex cases by leveraging their unique role capabilities. This section introduces the communicative framework for database diagnosis~\cite{agentverse,qian2023communicative}.

\begin{compactitem}
	\item
	\begin{sloppypar}
	\bfit{Agents.} In the communicative framework, agents can be undertaken by human beings or \llms. Humans can provide \llm agents with scenario requirements (e.g., business changes over the incoming period) and prior knowledge (e.g., historical anomalies). On the other hand, each \llm agent is dedicated to a distinct domain of functions. For example, we include three \llm agents in the initial implementation: (1) Chief DBA is responsible for collaboratively diagnosing and detecting root causes with other agents; (2) CPU Agent is specialized in CPU usage analysis and diagnosis, and (3) Memory Agent focuses on memory usage analysis and diagnosis. Each \llm agent can automatically invoke tool APIs to retrieve database statistics, extract external knowledge, and conduction optimizations. For instance, CPU Agent utilizes the monitoring tool {\it Prometheus} to check CPU usage metrics within specific time periods, and determine the root causes of high CPU usage by matching with extracted experience (Section~\ref{sec:prepare}). Note, if CPU/memory agents cannot report useful analysis, Chief DBA is responsible to detect other potential problems, such as those on the application side.
	\end{sloppypar}	

	\vspace{.5em}

	\item
	\bfit{Environment Settings.} We need to set a series of principles for the agents to efficiently communicate, such as {\it (1) Chat Order:} To avoid the mutual negative influence, we only allow one \llm agent to ``speak'' (i.e., appending the analysis results to the chat records to let other agents know) at a time. To ensure flexible chat (e.g., if an agent cannot detect anything useful, it should not speak), we rely on Chief DBA to decide which agent to speak in each iteration (diagnosis scheduling); {\it (2) Visibility:} By default, we assume the analysis results of agents can be seen by each other, i.e., within the same chat records. In the future, we can split agents into different groups, where each group is in charge of different database clusters/instances and they do not share the chat records; {\it (3) Selector} is vital to filter invalid analysis that may mislead the diagnosis directions; {\it (4) Updater} works to update agent memory based on the historical records.

\sbox{\blackball}{\tikz \fill[red] (0,0) circle (.5ex);}
\sbox{\greenball}{\tikz \fill[codegreen] (0,0) circle (.5ex);}

\begin{table*}[htbp]
	\centering
	\caption{Diagnosis performance of single root causes (\usebox{\blackball} : legal diagnosis results; \usebox{\greenball} : accurate diagnosis results).}
	\label{tbl:results}
	\vspace{-.25em}
	\begin{tabulary}{\linewidth}{|c|l|L|l|l|}
		\hline
		Type & \textbf{Root Cause} & \textbf{Description} & \textbf{LLM+Metrics} & \textbf{\oursys} \\
		\hline
		{Data Insert} & INSERT\_LARGE\_DATA & Long execution time for large data insertions & \usebox{\blackball} & \usebox{\blackball} \usebox{\greenball} \\
		\hline
		\multirow{6}{*}{\begin{tabular}[c]{@{}c@{}}Slow\\ Query\end{tabular}} & FETCH\_LARGE\_DATA & Fetching of large data volumes & \usebox{\blackball} \usebox{\greenball} & \usebox{\blackball} \usebox{\greenball} \\
		\cline{2-5}
		& REDUNDANT\_INDEX & Unnecessary and redundant indexes in tables & \usebox{\blackball} & \usebox{\blackball} \\
		\cline{2-5}
		& LACK\_STATISTIC\_INFO & Outdated statistical info affecting execution plan & \usebox{\blackball} & \usebox{\blackball} \usebox{\greenball} \\
		\cline{2-5}
		& MISSING\_INDEXES & Missing indexes causing performance issues & \usebox{\blackball} \usebox{\greenball} & \usebox{\blackball} \usebox{\greenball} \\
		\cline{2-5}
		& POOR\_JOIN\_PERFORMANCE & Poor performance of Join operators & \usebox{\blackball} & \usebox{\blackball} \usebox{\greenball} \\
		\cline{2-5}
		& CORRELATED\_SUBQUERY & Non-promotable subqueries in SQL & \usebox{\blackball} & \usebox{\blackball} \usebox{\greenball} \\
		\hline
		\multirow{4}{*}{\begin{tabular}[c]{@{}c@{}}Concurrent\\ Transaction\end{tabular}} & LOCK\_CONTENTION & Lock contention issues & \usebox{\blackball} & \usebox{\blackball} \\
		\cline{2-5}
		& WORKLOAD\_CONTENTION & Workload concentration affecting SQL execution & \usebox{\blackball} \usebox{\greenball} & \usebox{\blackball} \usebox{\greenball} \\
		\cline{2-5}
		& CPU\_CONTENTION & Severe external CPU resource contention & \usebox{\blackball} \usebox{\greenball}  & \usebox{\blackball} \usebox{\greenball} \\
		\cline{2-5}
		& IO\_CONTENTION & IO resource contention affecting SQL performance & \usebox{\blackball} & \usebox{\blackball} \usebox{\greenball} \\
		\hline
	\end{tabulary}
\end{table*}

	\vspace{.5em}
		
	\item
	\bfit{Chat Summary .} For a complex database problem, it requires agents dozens of iterations to give in-depth analysis, leading to extremely long chat records. Thus, it is vital to effectively summarize the critical information from chat records without exceeding the maximal length of \llm prompts. To the end, we progressively summarize the lines of a record used with tools, including inputs for certain tools and the results returned by these tools. Based on the current summary, it extracts the goals intended to be solved with each call to the tool, and forms a new summary, e.g.,
\end{compactitem}

\begin{tcolorbox}[enhanced,sharp corners,
	width={8.5cm},
	colback=white,
	borderline={0.3mm}{0.3mm}{white},
	boxsep=-.5mm]
    {\bf [Current summary]}\\
    - I know the start and end time of the anomaly.
    
    {\bf [New Record]}\\
    Thought: Now that I have the start and end time of the anomaly, I need to diagnose the causes of the anomaly\\
    Action: is\_abnormal\_metric\\
    Action Input: \{``start\_time'': 1684600070, ``end\_time'': 1684600074, ``metric\_name'': ``cpu\_usage''\}\\
    Observation: ``The metric is abnormal''
    
    {\bf [New summary]}\\
    - I know the start and end time of the anomaly.\\
    - {\it {\color{codegreen}I searched for is\_abnormal\_metric, and I now know that the CPU usage is abnormal.}}
\end{tcolorbox}

With this communicative framework and well-defined communication principles, the collaborative diagnosis process among human and \llm agents becomes more efficient (e.g., parallel diagnosis) and effective (e.g., chat records could trigger investigating of in-depth metric observation and root cause analysis). 


%% file: 06-experiment.tex

\section{Preliminary EXPERIMENT RESULTS}
\label{sec:experiments}

\vspace{.5em}

\begin{figure}[!t]
	\vspace{.5em}
	\centering
	\includegraphics[width=1.05\linewidth, trim={0 7.2cm 0 0},clip]{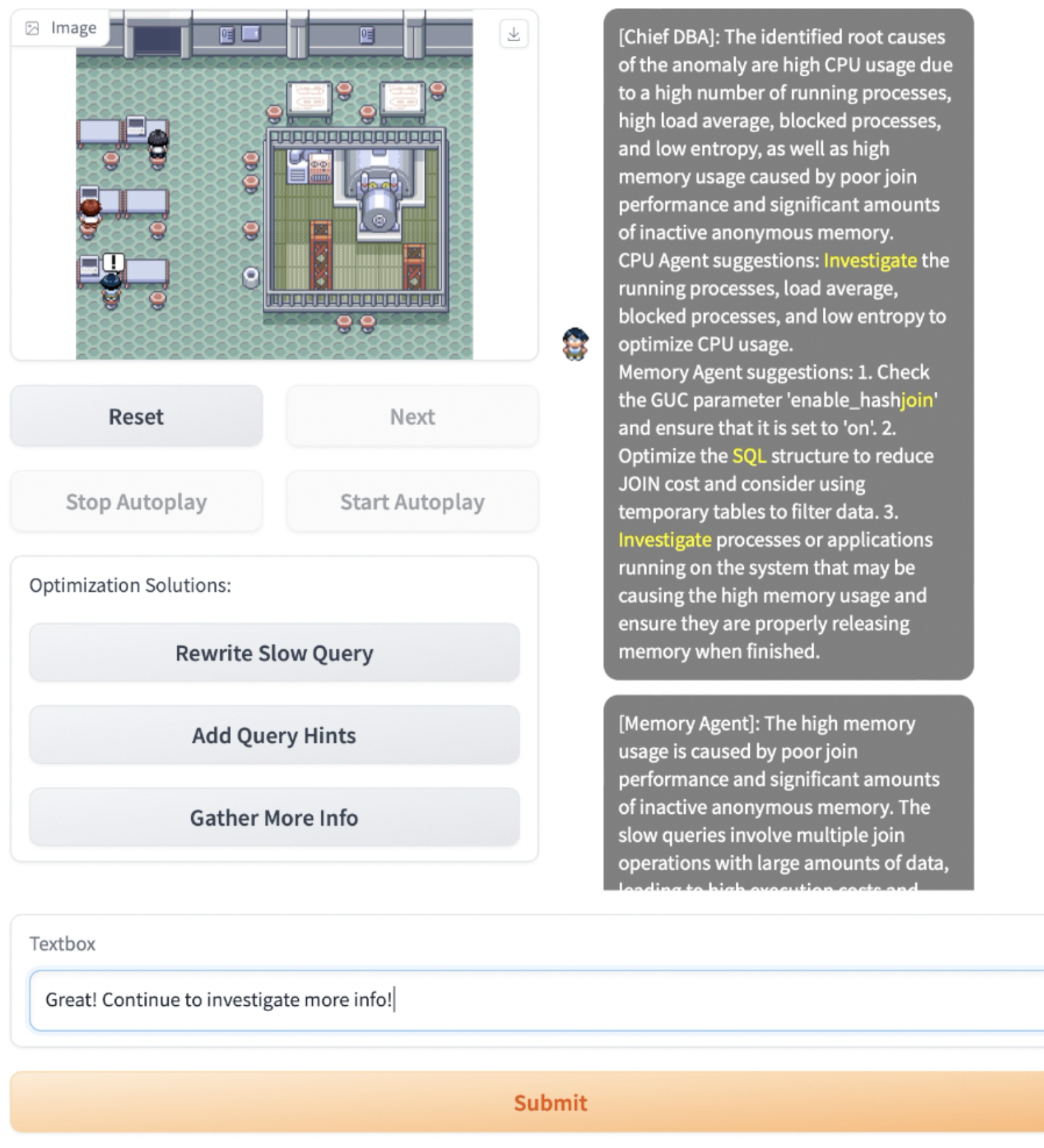}
	\vspace{-.5em}
	\caption{A basic demonstration of \oursys.}
	\label{fig:demo}
	\vspace{-.1em}
\end{figure}

\hi{Demonstration.} As illustrated in Figure~\ref{fig:demo}, {\it Chief DBA} monitors the status of the database to detect anomalies. Upon recognizing a new anomaly, {\it Chief DBA} notifies both the {\it Memory Agent} and {\it CPU Agent}. These agents independently assess the potential root causes and communicate their findings  (the root causes and recommended solutions) to the \textit{Chief DBA}. Subsequently, the \textit{Chief DBA} consolidates the diagnostic results for the user's convenience. In initial iterations, these agents generally gather limited information, and so they will continue for multiple iterations until the conclusion of {\it Chief DBA} is nearly certain or no further valuable information can be obtained. Additionally, during the diagnosis, users have the option to participate by offering instructions and feedback, such as verifying the effectiveness of a proposed optimization solution.

\hi{Diagnosis Performance Comparison.} We compare the performance of \oursys against a baseline, namely \texttt{llm+Metrics}. Both of the two methods are deployed with the OpenAI model GPT-4~\cite{openai} alongside metrics and views from PostgreSQL and Prometheus. The evaluation focuses on basic single-cause problems as detailed in Table~\ref{tbl:results}. Besides, we also offer a multi-cause diagnosis example presented in the \emph{Appendix-B}.

Preliminary results indicate that {\it \llm+Metrics} and \oursys can achieve a high \emph{legality rate} (producing valid responses to specific database issues). However, it is a ``dangerous behavior'' for  {\it \llm+Metrics}, which actually has very low {\it success rate (infrequent provision of the correct causes)}. In contrast, \oursys achieves both high legal rate and success rate. The reasons are three-fold.

First, {\it \llm+Metrics} conducts very basic reasoning and often misses key causes. For example,  for the {\it INSERT\_LARGE\_DATA} case, {\it \llm+Metrics} only finds ``high number of running processes'' with the {\it node\_procs\_running} metric, and stops early. In contrast, \oursys not only finds the high concurrency problem, but analyze the operation statistics in the database process and identifies {\it ``high memory usage due to heavy use of UPDATE and INSERT operations on xxx tables''} by looking up the {\it pg\_stat\_statements} view.

\begin{sloppypar}
Second, {\it \llm+Metrics} often ``makes up'' reasons without substantial knowledge evidence. For example, for the {\it CORRELATED\_SUBQUERY} case, {\it \llm+Metrics} observes SORT operations in logged queries, and incorrectly attributes the cause to ``frequent reading and sorting of large amount of data'', thereby ending the diagnostic process. Instead, \oursys cross-references with the query optimization knowledge, and then finds the correlated-subquery structure might be the performance bottleneck, with additional extracted information like estimated operation costs.
\end{sloppypar}

Third, {\it \llm+Metrics} meet trouble in deriving appropriate solutions. {\it \llm+Metrics} often gives very generic optimization solutions (e.g., ``resolve resource contention issues''), which are useless in practice. Instead, leveraging its \emph{tool retrieval} component, \oursys can learn to give specific optimization advice (e.g., invoking query transformation rules, adjusting the work\_mem parameter) or gather more insightful information (e.g., ``calculate the total cost of the plan and check whether the cost rate of the sort or hash operators exceeds the cost rate threshold'').

This evaluation reveals the potential of \oursys in going beyond mere anomaly detection to root cause analysis and provision of actionable suggestions. Despite these advancements, from the basic deployment of \oursys, there are still some unresolved challenges. First, it is tricky to share the maintenance experience (e.g., varying metric and view names) across different database products. Second, it is labor-intensive to adequately prepare extensive number of anomaly-diagnosis data, which is essential to fine-tune and direct less-capable \llms (e.g., those smaller than 10B) to understand the complex database knowledge and apply in maintenance.

%% file: 09-conclusion.tex
\section{Conclusion}
\label{sec:conclusion}

In this paper, we propose a vision of \oursys, an \llm-based database administrator that can continuously acquire database maintenance experience from textual sources, and provide reasonable, well-founded, in-time diagnosis and optimization advice for target databases. {\it We will continue to complete and improve this work with our collaborators.}



%% file: appendix.tex
\section{Appendix - Prompts}
\label{sec:prompts}
\centering
\includegraphics[width=.87\textwidth]{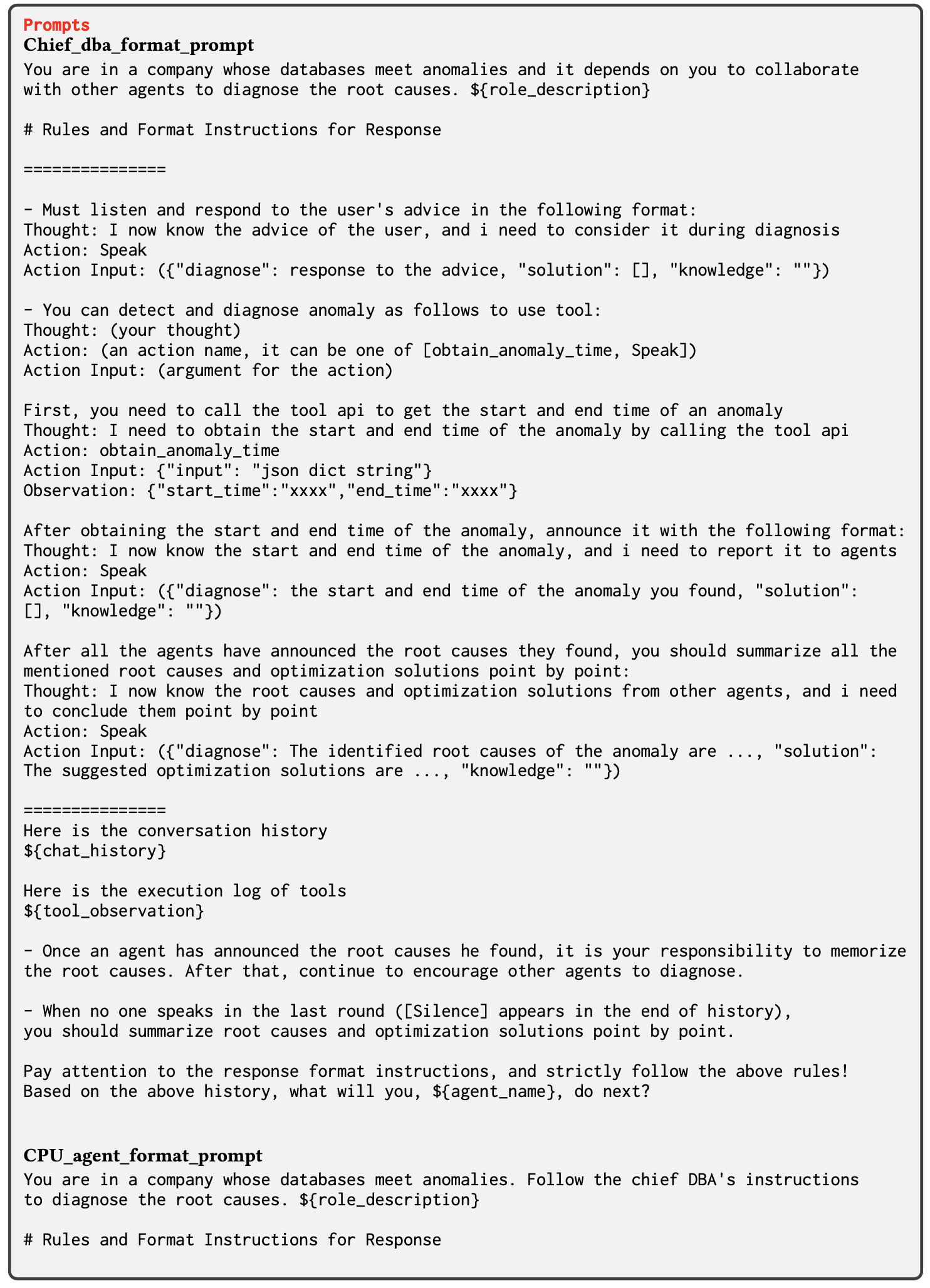}


\centering
\includegraphics[width=.87\textwidth]{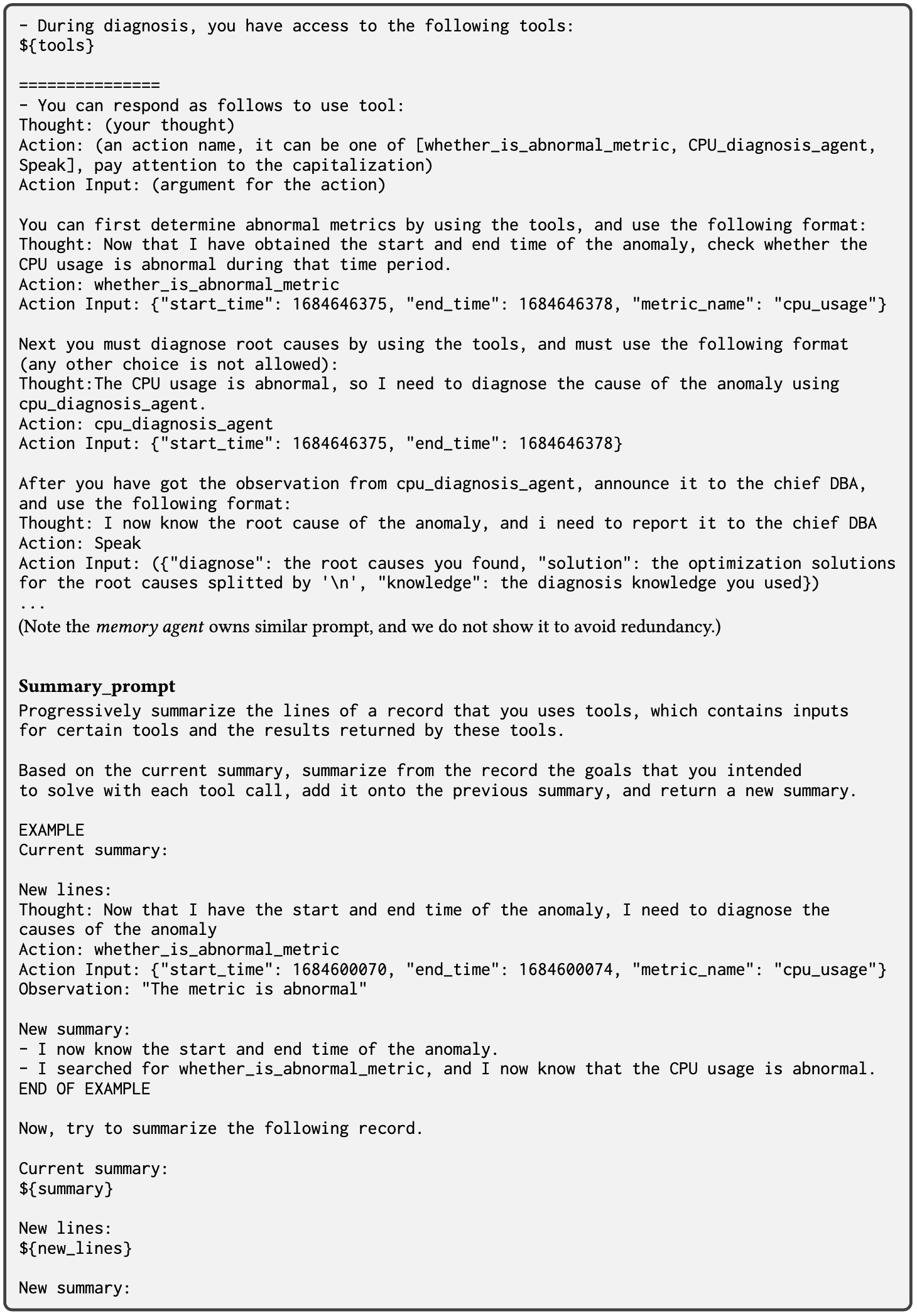}

\section{Appendix - Test Cases}
\label{sec:tests}

\centering
\includegraphics[width=.87\textwidth]{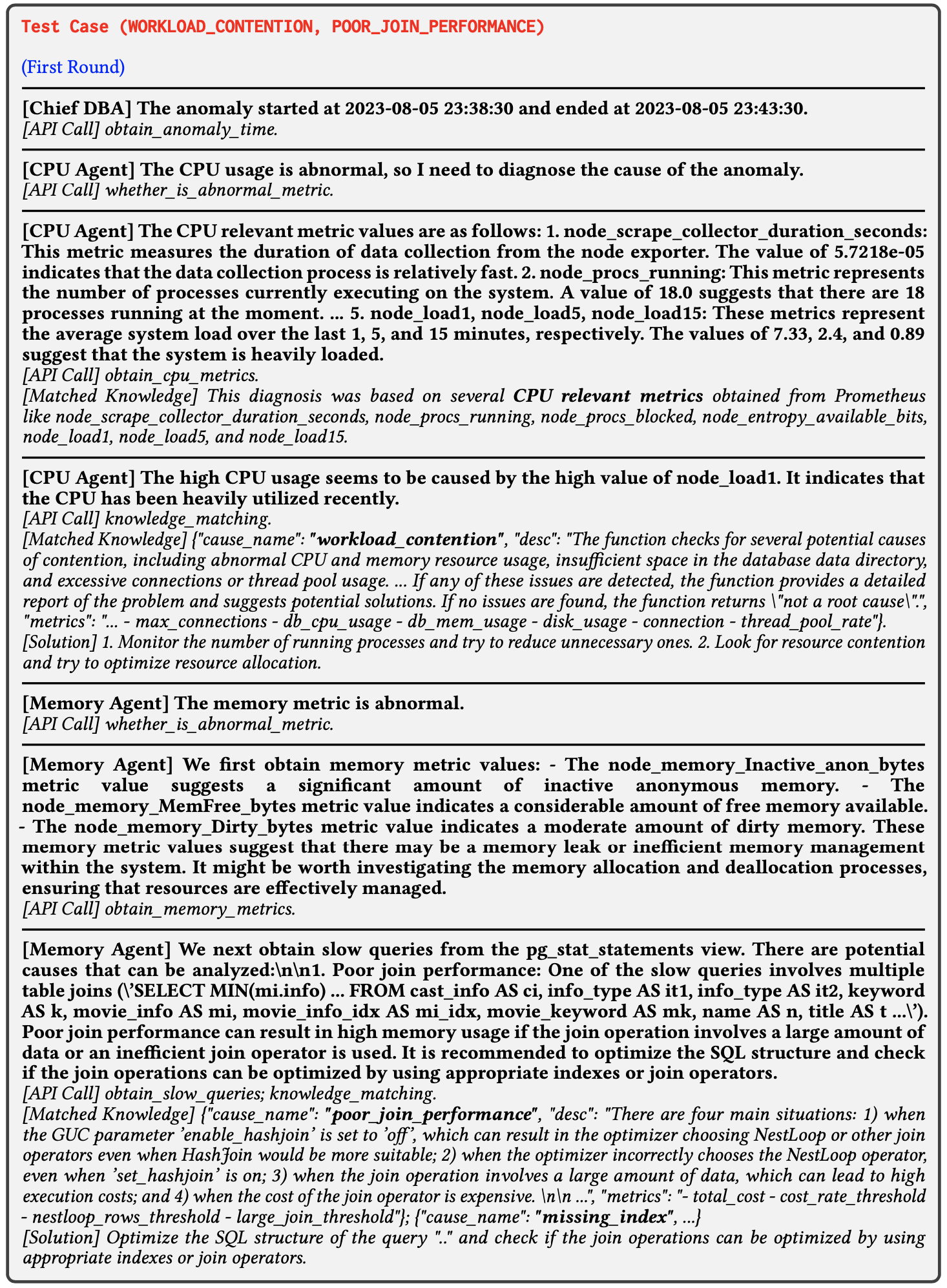}

\centering
\includegraphics[width=.87\textwidth]{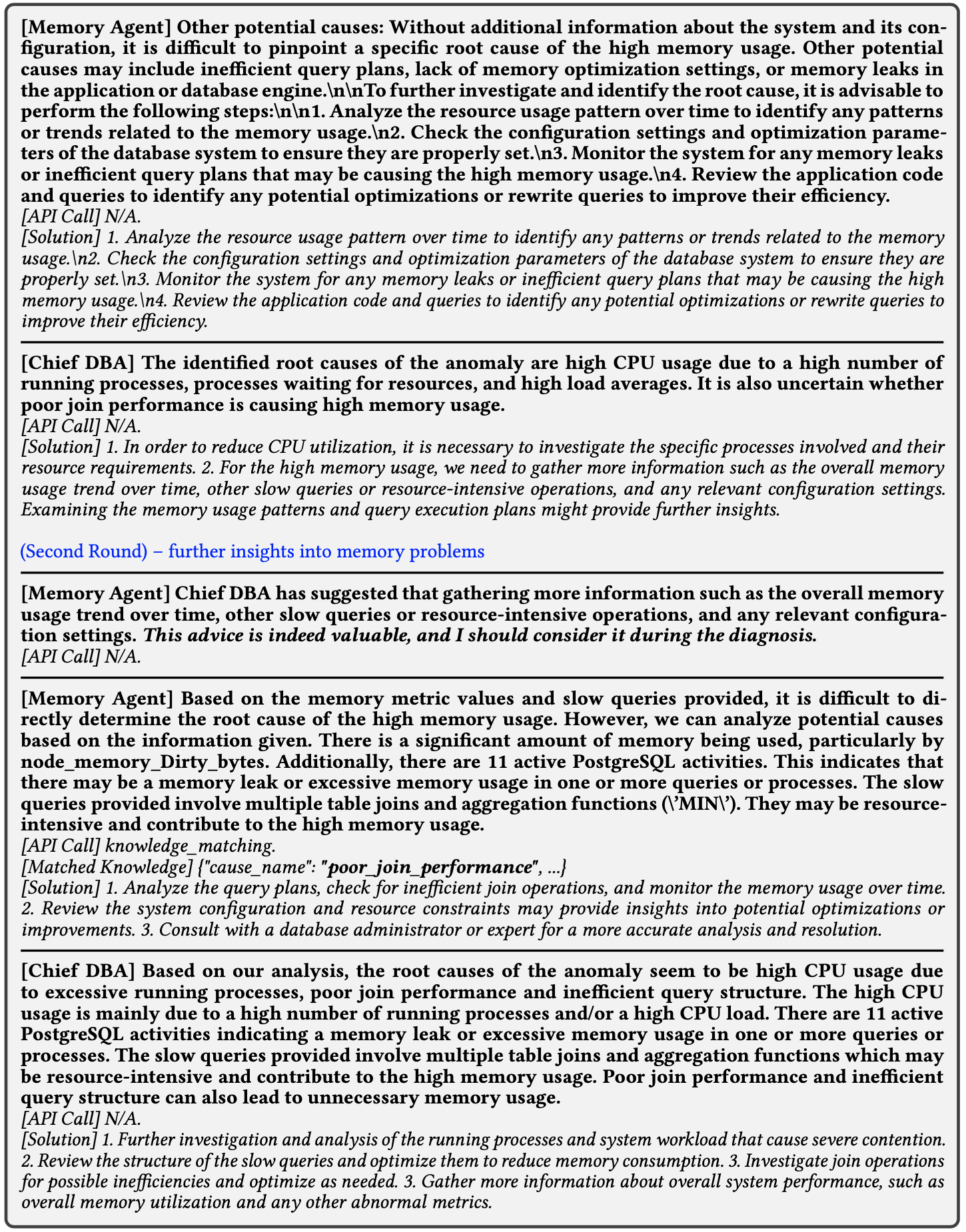}

%% file: main.bbl

\begin{thebibliography}{27}


\ifx \showCODEN    \undefined \def \showCODEN     #1{\unskip}     \fi
\ifx \showDOI      \undefined \def \showDOI       #1{#1}\fi
\ifx \showISBNx    \undefined \def \showISBNx     #1{\unskip}     \fi
\ifx \showISBNxiii \undefined \def \showISBNxiii  #1{\unskip}     \fi
\ifx \showISSN     \undefined \def \showISSN      #1{\unskip}     \fi
\ifx \showLCCN     \undefined \def \showLCCN      #1{\unskip}     \fi
\ifx \shownote     \undefined \def \shownote      #1{#1}          \fi
\ifx \showarticletitle \undefined \def \showarticletitle #1{#1}   \fi
\ifx \showURL      \undefined \def \showURL       {\relax}        \fi
\providecommand\bibfield[2]{#2}
\providecommand\bibinfo[2]{#2}
\providecommand\natexlab[1]{#1}
\providecommand\showeprint[2][]{arXiv:#2}

\bibitem[\protect\citeauthoryear{??}{age}{[n.d.]}]%
        {agentverse}
 \bibinfo{year}{[n.d.]}\natexlab{}.
\newblock \showarticletitle{https://github.com/OpenBMB/AgentVerse}.
\newblock
\newblock
\shownote{Last accessed on 2023-8.}


\bibitem[\protect\citeauthoryear{??}{ope}{[n.d.]}]%
        {openai}
 \bibinfo{year}{[n.d.]}\natexlab{}.
\newblock \showarticletitle{https://openai.com/}.
\newblock
\newblock
\shownote{Last accessed on 2023-8.}


\bibitem[\protect\citeauthoryear{Chaudhuri and Narasayya}{Chaudhuri and
  Narasayya}{1997}]%
        {AutoAdmin}
\bibfield{author}{\bibinfo{person}{Surajit Chaudhuri} {and}
  \bibinfo{person}{Vivek~R. Narasayya}.} \bibinfo{year}{1997}\natexlab{}.
\newblock \showarticletitle{An Efficient Cost-Driven Index Selection Tool for
  Microsoft {SQL} Server}. In \bibinfo{booktitle}{\emph{{VLDB}}}.
  \bibinfo{pages}{146--155}.
\newblock


\bibitem[\protect\citeauthoryear{Dias, Ramacher, Shaft, Venkataramani, and
  Wood}{Dias et~al\mbox{.}}{2005}]%
        {DBLP:conf/cidr/DiasRSVW05}
\bibfield{author}{\bibinfo{person}{Karl Dias}, \bibinfo{person}{Mark Ramacher},
  \bibinfo{person}{Uri Shaft}, \bibinfo{person}{Venkateshwaran Venkataramani},
  {and} \bibinfo{person}{Graham Wood}.} \bibinfo{year}{2005}\natexlab{}.
\newblock \showarticletitle{Automatic Performance Diagnosis and Tuning in
  Oracle}. In \bibinfo{booktitle}{\emph{Second Biennial Conference on
  Innovative Data Systems Research, {CIDR} 2005, Asilomar, CA, USA, January
  4-7, 2005, Online Proceedings}}. \bibinfo{publisher}{www.cidrdb.org},
  \bibinfo{pages}{84--94}.
\newblock
\urldef\tempurl%
\url{http://cidrdb.org/cidr2005/papers/P07.pdf}
\showURL{%
\tempurl}


\bibitem[\protect\citeauthoryear{Huang, Wang, Zhang, Tu, Li, and Cui}{Huang
  et~al\mbox{.}}{2023}]%
        {DBLP:journals/pacmmod/HuangWZTL023}
\bibfield{author}{\bibinfo{person}{Shiyue Huang}, \bibinfo{person}{Ziwei Wang},
  \bibinfo{person}{Xinyi Zhang}, \bibinfo{person}{Yaofeng Tu},
  \bibinfo{person}{Zhongliang Li}, {and} \bibinfo{person}{Bin Cui}.}
  \bibinfo{year}{2023}\natexlab{}.
\newblock \showarticletitle{{DBPA:} {A} Benchmark for Transactional Database
  Performance Anomalies}.
\newblock \bibinfo{journal}{\emph{Proc. {ACM} Manag. Data}}
  \bibinfo{volume}{1}, \bibinfo{number}{1} (\bibinfo{year}{2023}),
  \bibinfo{pages}{72:1--72:26}.
\newblock
\urldef\tempurl%
\url{https://doi.org/10.1145/3588926}
\showDOI{\tempurl}


\bibitem[\protect\citeauthoryear{Kalmegh, Babu, and Roy}{Kalmegh
  et~al\mbox{.}}{2019}]%
        {DBLP:conf/sigmod/KalmeghBR19}
\bibfield{author}{\bibinfo{person}{Prajakta Kalmegh}, \bibinfo{person}{Shivnath
  Babu}, {and} \bibinfo{person}{Sudeepa Roy}.} \bibinfo{year}{2019}\natexlab{}.
\newblock \showarticletitle{iQCAR: inter-Query Contention Analyzer for Data
  Analytics Frameworks}. In \bibinfo{booktitle}{\emph{Proceedings of the 2019
  International Conference on Management of Data, {SIGMOD} Conference 2019,
  Amsterdam, The Netherlands, June 30 - July 5, 2019}},
  \bibfield{editor}{\bibinfo{person}{Peter~A. Boncz}, \bibinfo{person}{Stefan
  Manegold}, \bibinfo{person}{Anastasia Ailamaki}, \bibinfo{person}{Amol
  Deshpande}, {and} \bibinfo{person}{Tim Kraska}} (Eds.).
  \bibinfo{publisher}{{ACM}}, \bibinfo{pages}{918--935}.
\newblock
\urldef\tempurl%
\url{https://doi.org/10.1145/3299869.3319904}
\showDOI{\tempurl}


\bibitem[\protect\citeauthoryear{Kossmann, Kastius, and Schlosser}{Kossmann
  et~al\mbox{.}}{2022}]%
        {SWIRL}
\bibfield{author}{\bibinfo{person}{Jan Kossmann}, \bibinfo{person}{Alexander
  Kastius}, {and} \bibinfo{person}{Rainer Schlosser}.}
  \bibinfo{year}{2022}\natexlab{}.
\newblock \showarticletitle{{SWIRL:} Selection of Workload-aware Indexes using
  Reinforcement Learning}. In \bibinfo{booktitle}{\emph{{EDBT}}}.
  \bibinfo{pages}{2:155--2:168}.
\newblock


\bibitem[\protect\citeauthoryear{Lan, Bao, and Peng}{Lan et~al\mbox{.}}{2020}]%
        {CIKM20}
\bibfield{author}{\bibinfo{person}{Hai Lan}, \bibinfo{person}{Zhifeng Bao},
  {and} \bibinfo{person}{Yuwei Peng}.} \bibinfo{year}{2020}\natexlab{}.
\newblock \showarticletitle{An Index Advisor Using Deep Reinforcement
  Learning}. In \bibinfo{booktitle}{\emph{{CIKM}}}.
  \bibinfo{pages}{2105--2108}.
\newblock


\bibitem[\protect\citeauthoryear{Licks, Couto, de~F{\'{a}}tima~Miehe, Paris,
  Ruiz, and Meneguzzi}{Licks et~al\mbox{.}}{2020}]%
        {SmartIX}
\bibfield{author}{\bibinfo{person}{Gabriel~Paludo Licks},
  \bibinfo{person}{J{\'{u}}lia Mara~Colleoni Couto}, \bibinfo{person}{Priscilla
  de F{\'{a}}tima~Miehe}, \bibinfo{person}{Renata~De Paris},
  \bibinfo{person}{Duncan Dubugras~A. Ruiz}, {and} \bibinfo{person}{Felipe
  Meneguzzi}.} \bibinfo{year}{2020}\natexlab{}.
\newblock \showarticletitle{SmartIX: {A} database indexing agent based on
  reinforcement learning}.
\newblock \bibinfo{journal}{\emph{Appl. Intell.}} \bibinfo{volume}{50},
  \bibinfo{number}{8} (\bibinfo{year}{2020}), \bibinfo{pages}{2575--2588}.
\newblock


\bibitem[\protect\citeauthoryear{Liu, Zhang, Sun, Meng, Yang, and Pei}{Liu
  et~al\mbox{.}}{2020}]%
        {DBLP:conf/ipccc/LiuZSMYP20}
\bibfield{author}{\bibinfo{person}{Ping Liu}, \bibinfo{person}{Shenglin Zhang},
  \bibinfo{person}{Yongqian Sun}, \bibinfo{person}{Yuan Meng},
  \bibinfo{person}{Jiahai Yang}, {and} \bibinfo{person}{Dan Pei}.}
  \bibinfo{year}{2020}\natexlab{}.
\newblock \showarticletitle{FluxInfer: Automatic Diagnosis of Performance
  Anomaly for Online Database System}. In \bibinfo{booktitle}{\emph{39th {IEEE}
  International Performance Computing and Communications Conference, {IPCCC}
  2020, Austin, TX, USA, November 6-8, 2020}}. \bibinfo{publisher}{{IEEE}},
  \bibinfo{pages}{1--8}.
\newblock
\urldef\tempurl%
\url{https://doi.org/10.1109/IPCCC50635.2020.9391550}
\showDOI{\tempurl}


\bibitem[\protect\citeauthoryear{Liu, Yin, Zhao, Ge, Chen, Gao, Li, Wang,
  Liang, Tan, and Li}{Liu et~al\mbox{.}}{2022}]%
        {DBLP:conf/icde/LiuYZGCGLWLTL22}
\bibfield{author}{\bibinfo{person}{Xiaoze Liu}, \bibinfo{person}{Zheng Yin},
  \bibinfo{person}{Chao Zhao}, \bibinfo{person}{Congcong Ge},
  \bibinfo{person}{Lu Chen}, \bibinfo{person}{Yunjun Gao},
  \bibinfo{person}{Dimeng Li}, \bibinfo{person}{Ziting Wang},
  \bibinfo{person}{Gaozhong Liang}, \bibinfo{person}{Jian Tan}, {and}
  \bibinfo{person}{Feifei Li}.} \bibinfo{year}{2022}\natexlab{}.
\newblock \showarticletitle{PinSQL: Pinpoint Root Cause SQLs to Resolve
  Performance Issues in Cloud Databases}. In \bibinfo{booktitle}{\emph{38th
  {IEEE} International Conference on Data Engineering, {ICDE} 2022, Kuala
  Lumpur, Malaysia, May 9-12, 2022}}. \bibinfo{publisher}{{IEEE}},
  \bibinfo{pages}{2549--2561}.
\newblock
\urldef\tempurl%
\url{https://doi.org/10.1109/ICDE53745.2022.00236}
\showDOI{\tempurl}


\bibitem[\protect\citeauthoryear{Lu, Xie, Li, Li, Nie, Zhao, Yu, Zhang, Sui,
  Zhu, and Pei}{Lu et~al\mbox{.}}{2022}]%
        {DBLP:conf/ccgrid/LuXL0NZYZSZP22}
\bibfield{author}{\bibinfo{person}{Xianglin Lu}, \bibinfo{person}{Zhe Xie},
  \bibinfo{person}{Zeyan Li}, \bibinfo{person}{Mingjie Li},
  \bibinfo{person}{Xiaohui Nie}, \bibinfo{person}{Nengwen Zhao},
  \bibinfo{person}{Qingyang Yu}, \bibinfo{person}{Shenglin Zhang},
  \bibinfo{person}{Kaixin Sui}, \bibinfo{person}{Lin Zhu}, {and}
  \bibinfo{person}{Dan Pei}.} \bibinfo{year}{2022}\natexlab{}.
\newblock \showarticletitle{Generic and Robust Performance Diagnosis via Causal
  Inference for {OLTP} Database Systems}. In \bibinfo{booktitle}{\emph{22nd
  {IEEE} International Symposium on Cluster, Cloud and Internet Computing,
  CCGrid 2022, Taormina, Italy, May 16-19, 2022}}. \bibinfo{publisher}{{IEEE}},
  \bibinfo{pages}{655--664}.
\newblock
\urldef\tempurl%
\url{https://doi.org/10.1109/CCGrid54584.2022.00075}
\showDOI{\tempurl}


\bibitem[\protect\citeauthoryear{Ma, Yin, Zhang, and et~al}{Ma
  et~al\mbox{.}}{2020}]%
        {DBLP:journals/pvldb/MaYZWZJHLLQLCP20}
\bibfield{author}{\bibinfo{person}{Minghua Ma}, \bibinfo{person}{Zheng Yin},
  \bibinfo{person}{Shenglin Zhang}, {and} \bibinfo{person}{et al}.}
  \bibinfo{year}{2020}\natexlab{}.
\newblock \showarticletitle{Diagnosing Root Causes of Intermittent Slow Queries
  in Large-Scale Cloud Databases}.
\newblock \bibinfo{journal}{\emph{Proc. {VLDB} Endow.}} \bibinfo{volume}{13},
  \bibinfo{number}{8} (\bibinfo{year}{2020}), \bibinfo{pages}{1176--1189}.
\newblock
\urldef\tempurl%
\url{https://doi.org/10.14778/3389133.3389136}
\showDOI{\tempurl}


\bibitem[\protect\citeauthoryear{Ma, Zhang, and Zhu}{Ma et~al\mbox{.}}{2023}]%
        {DBLP:journals/corr/abs-2307-03762}
\bibfield{author}{\bibinfo{person}{Yuxi Ma}, \bibinfo{person}{Chi Zhang}, {and}
  \bibinfo{person}{Song{-}Chun Zhu}.} \bibinfo{year}{2023}\natexlab{}.
\newblock \showarticletitle{Brain in a Vat: On Missing Pieces Towards
  Artificial General Intelligence in Large Language Models}.
\newblock \bibinfo{journal}{\emph{CoRR}}  \bibinfo{volume}{abs/2307.03762}
  (\bibinfo{year}{2023}).
\newblock
\urldef\tempurl%
\url{https://doi.org/10.48550/arXiv.2307.03762}
\showDOI{\tempurl}
\showeprint[arXiv]{2307.03762}


\bibitem[\protect\citeauthoryear{Perera, Oetomo, Rubinstein, and
  Borovica{-}Gajic}{Perera et~al\mbox{.}}{2021}]%
        {DBABandits}
\bibfield{author}{\bibinfo{person}{R.~Malinga Perera}, \bibinfo{person}{Bastian
  Oetomo}, \bibinfo{person}{Benjamin I.~P. Rubinstein}, {and}
  \bibinfo{person}{Renata Borovica{-}Gajic}.} \bibinfo{year}{2021}\natexlab{}.
\newblock \showarticletitle{{DBA} bandits: Self-driving index tuning under
  ad-hoc, analytical workloads with safety guarantees}. In
  \bibinfo{booktitle}{\emph{{ICDE}}}. \bibinfo{pages}{600--611}.
\newblock


\bibitem[\protect\citeauthoryear{Qian, Cong, Yang, Chen, Su, and et~al}{Qian
  et~al\mbox{.}}{2023}]%
        {qian2023communicative}
\bibfield{author}{\bibinfo{person}{Chen Qian}, \bibinfo{person}{Xin Cong},
  \bibinfo{person}{Cheng Yang}, \bibinfo{person}{Weize Chen},
  \bibinfo{person}{Yusheng Su}, {and} \bibinfo{person}{et al}.}
  \bibinfo{year}{2023}\natexlab{}.
\newblock \showarticletitle{Communicative Agents for Software Development}.
\newblock \bibinfo{journal}{\emph{arXiv preprint arXiv:2307.07924}}
  (\bibinfo{year}{2023}).
\newblock


\bibitem[\protect\citeauthoryear{Qin, Hu, Lin, and et~al}{Qin
  et~al\mbox{.}}{2023a}]%
        {qin2023tool}
\bibfield{author}{\bibinfo{person}{Yujia Qin}, \bibinfo{person}{Shengding Hu},
  \bibinfo{person}{Yankai Lin}, {and} \bibinfo{person}{et al}.}
  \bibinfo{year}{2023}\natexlab{a}.
\newblock \showarticletitle{Tool learning with foundation models}.
\newblock \bibinfo{journal}{\emph{arXiv preprint arXiv:2304.08354}}
  (\bibinfo{year}{2023}).
\newblock


\bibitem[\protect\citeauthoryear{Qin, Liang, Ye, Zhu, Yan, Lu, Lin, Cong, Tang,
  Qian, Zhao, Tian, Xie, Zhou, Gerstein, Li, Liu, and Sun}{Qin
  et~al\mbox{.}}{2023b}]%
        {qin2023toolllm}
\bibfield{author}{\bibinfo{person}{Yujia Qin}, \bibinfo{person}{Shihao Liang},
  \bibinfo{person}{Yining Ye}, \bibinfo{person}{Kunlun Zhu},
  \bibinfo{person}{Lan Yan}, \bibinfo{person}{Yaxi Lu}, \bibinfo{person}{Yankai
  Lin}, \bibinfo{person}{Xin Cong}, \bibinfo{person}{Xiangru Tang},
  \bibinfo{person}{Bill Qian}, \bibinfo{person}{Sihan Zhao},
  \bibinfo{person}{Runchu Tian}, \bibinfo{person}{Ruobing Xie},
  \bibinfo{person}{Jie Zhou}, \bibinfo{person}{Mark Gerstein},
  \bibinfo{person}{Dahai Li}, \bibinfo{person}{Zhiyuan Liu}, {and}
  \bibinfo{person}{Maosong Sun}.} \bibinfo{year}{2023}\natexlab{b}.
\newblock \bibinfo{title}{ToolLLM: Facilitating Large Language Models to Master
  16000+ Real-world APIs}.
\newblock
\newblock
\showeprint[arxiv]{cs.AI/2307.16789}


\bibitem[\protect\citeauthoryear{Robertson, Zaragoza, et~al\mbox{.}}{Robertson
  et~al\mbox{.}}{2009}]%
        {robertson2009probabilistic}
\bibfield{author}{\bibinfo{person}{Stephen Robertson}, \bibinfo{person}{Hugo
  Zaragoza}, {et~al\mbox{.}}} \bibinfo{year}{2009}\natexlab{}.
\newblock \showarticletitle{The probabilistic relevance framework: BM25 and
  beyond}.
\newblock \bibinfo{journal}{\emph{Foundations and Trends{\textregistered} in
  Information Retrieval}} \bibinfo{volume}{3}, \bibinfo{number}{4}
  (\bibinfo{year}{2009}), \bibinfo{pages}{333--389}.
\newblock


\bibitem[\protect\citeauthoryear{Turnbull}{Turnbull}{2018}]%
        {turnbull2018monitoring}
\bibfield{author}{\bibinfo{person}{James Turnbull}.}
  \bibinfo{year}{2018}\natexlab{}.
\newblock \bibinfo{booktitle}{\emph{Monitoring with Prometheus}}.
\newblock \bibinfo{publisher}{Turnbull Press}.
\newblock


\bibitem[\protect\citeauthoryear{Valentin, Zuliani, Zilio, Lohman, and
  Skelley}{Valentin et~al\mbox{.}}{2000}]%
        {DB2Advis}
\bibfield{author}{\bibinfo{person}{Gary Valentin}, \bibinfo{person}{Michael
  Zuliani}, \bibinfo{person}{Daniel~C. Zilio}, \bibinfo{person}{Guy~M. Lohman},
  {and} \bibinfo{person}{Alan Skelley}.} \bibinfo{year}{2000}\natexlab{}.
\newblock \showarticletitle{{DB2} Advisor: An Optimizer Smart Enough to
  Recommend Its Own Indexes}. In \bibinfo{booktitle}{\emph{{ICDE}}}.
  \bibinfo{pages}{101--110}.
\newblock


\bibitem[\protect\citeauthoryear{Whang}{Whang}{1987}]%
        {Drop}
\bibfield{author}{\bibinfo{person}{Kyu{-}Young Whang}.}
  \bibinfo{year}{1987}\natexlab{}.
\newblock \showarticletitle{Index Selection in Relational Databases}.
\newblock \bibinfo{journal}{\emph{Foundations of Data Organization}}
  (\bibinfo{year}{1987}), \bibinfo{pages}{487--500}.
\newblock


\bibitem[\protect\citeauthoryear{Wu, Wang, Siddiqui, Wang, Narasayya,
  Chaudhuri, and Bernstein}{Wu et~al\mbox{.}}{2022}]%
        {BudgetMCTS}
\bibfield{author}{\bibinfo{person}{Wentao Wu}, \bibinfo{person}{Chi Wang},
  \bibinfo{person}{Tarique Siddiqui}, \bibinfo{person}{Junxiong Wang},
  \bibinfo{person}{Vivek~R. Narasayya}, \bibinfo{person}{Surajit Chaudhuri},
  {and} \bibinfo{person}{Philip~A. Bernstein}.}
  \bibinfo{year}{2022}\natexlab{}.
\newblock \showarticletitle{Budget-aware Index Tuning with Reinforcement
  Learning}. In \bibinfo{booktitle}{\emph{{SIGMOD} Conference}}.
  \bibinfo{pages}{1528--1541}.
\newblock


\bibitem[\protect\citeauthoryear{Yoon, Niu, and Mozafari}{Yoon
  et~al\mbox{.}}{2016}]%
        {DBLP:conf/sigmod/YoonNM16}
\bibfield{author}{\bibinfo{person}{Dong~Young Yoon}, \bibinfo{person}{Ning
  Niu}, {and} \bibinfo{person}{Barzan Mozafari}.}
  \bibinfo{year}{2016}\natexlab{}.
\newblock \showarticletitle{DBSherlock: {A} Performance Diagnostic Tool for
  Transactional Databases}. In \bibinfo{booktitle}{\emph{Proceedings of the
  2016 International Conference on Management of Data, {SIGMOD} Conference
  2016, San Francisco, CA, USA, June 26 - July 01, 2016}},
  \bibfield{editor}{\bibinfo{person}{Fatma {\"{O}}zcan},
  \bibinfo{person}{Georgia Koutrika}, {and} \bibinfo{person}{Sam Madden}}
  (Eds.). \bibinfo{publisher}{{ACM}}, \bibinfo{pages}{1599--1614}.
\newblock
\urldef\tempurl%
\url{https://doi.org/10.1145/2882903.2915218}
\showDOI{\tempurl}


\bibitem[\protect\citeauthoryear{Zhou, Chai, Li, and Sun}{Zhou
  et~al\mbox{.}}{2020}]%
        {zhou2020database}
\bibfield{author}{\bibinfo{person}{Xuanhe Zhou}, \bibinfo{person}{Chengliang
  Chai}, \bibinfo{person}{Guoliang Li}, {and} \bibinfo{person}{Ji Sun}.}
  \bibinfo{year}{2020}\natexlab{}.
\newblock \showarticletitle{Database meets artificial intelligence: A survey}.
\newblock \bibinfo{journal}{\emph{IEEE Transactions on Knowledge and Data
  Engineering}} \bibinfo{volume}{34}, \bibinfo{number}{3}
  (\bibinfo{year}{2020}), \bibinfo{pages}{1096--1116}.
\newblock


\bibitem[\protect\citeauthoryear{Zhou, Liu, Li, Jin, Li, Wang, and Feng}{Zhou
  et~al\mbox{.}}{2022a}]%
        {AutoIndex}
\bibfield{author}{\bibinfo{person}{Xuanhe Zhou}, \bibinfo{person}{Luyang Liu},
  \bibinfo{person}{Wenbo Li}, \bibinfo{person}{Lianyuan Jin},
  \bibinfo{person}{Shifu Li}, \bibinfo{person}{Tianqing Wang}, {and}
  \bibinfo{person}{Jianhua Feng}.} \bibinfo{year}{2022}\natexlab{a}.
\newblock \showarticletitle{AutoIndex: An Incremental Index Management System
  for Dynamic Workloads}. In \bibinfo{booktitle}{\emph{{ICDE}}}.
  \bibinfo{pages}{2196--2208}.
\newblock


\bibitem[\protect\citeauthoryear{Zhou, Muresanu, Han, Paster, Pitis, Chan, and
  Ba}{Zhou et~al\mbox{.}}{2022b}]%
        {Zhou2022a}
\bibfield{author}{\bibinfo{person}{Yongchao Zhou}, \bibinfo{person}{Andrei~Ioan
  Muresanu}, \bibinfo{person}{Ziwen Han}, \bibinfo{person}{Keiran Paster},
  \bibinfo{person}{Silviu Pitis}, \bibinfo{person}{Harris Chan}, {and}
  \bibinfo{person}{Jimmy Ba}.} \bibinfo{year}{2022}\natexlab{b}.
\newblock \showarticletitle{{Large Language Models Are Human-Level Prompt
  Engineers}}.
\newblock  (\bibinfo{year}{2022}).
\newblock
\showeprint[arxiv]{2211.01910}
\urldef\tempurl%
\url{http://arxiv.org/abs/2211.01910}
\showURL{%
\tempurl}


\end{thebibliography}
